\documentclass[prb,twocolumn,superscriptaddress,showpacs,
amsmath,amssymb]{revtex4}

\usepackage{graphicx} 
\usepackage{dcolumn} 

\def\b0{{\bf 0}}

\begin{document}

\title{Quantum interface unbinding transitions}

\author{P.~Jakubczyk}
\email{pjak@fuw.edu.pl}
\affiliation{Institute of Theoretical Physics, University of Warsaw, 
 Ho\.za 69, 00-681 Warsaw, Poland}

\date{\today}

\begin{abstract}
We consider interfacial phenomena accompanying bulk quantum phase transitions in presence of surface fields. On general grounds we argue that the surface contribution to the system free 
energy involves a line of singularities characteristic of an interfacial phase transition, occurring below the bulk transition temperature $T_c$ down to $T=0$. 
This implies the occurrence of an interfacial quantum critical regime extending into finite temperatures and located within the portion of the phase diagram where the bulk is ordered. 
Even in situations, where the bulk order sets in discontinuously at $T=0$, the system's behavior at the boundary may be controlled 
by a divergent length scale if the tricritical temperature is sufficiently low. 
Relying on an effective interfacial model we compute the surface phase diagram in bulk spatial dimensionality $d\geq 2$ and extract the values of the exponents describing the interfacial singularities in $d\geq 3$. 
     
\end{abstract}

\maketitle

\section{Introduction}
The theory of interacting quantum systems undergoing bulk ground state phase transitions is presently an arena of fascinating developments strongly linked to parallelly progressing experimental work.~\cite{Sachdev_book, 
Science, Nature, Stat_Sol, Belitz_05, Loehneysen_07}
Most of the theoretical interest is concerned with bulk properties of homogeneous systems. Realistic experimental samples are however finite and inhomogeneous at the boundaries. The issues concerning the 
boundaries' influence on bulk critical behavior and the character of interfacial singularities at bulk criticality were widely investigated for classical phase transitions.~\cite{Binder_DL} So were interface unbinding transitions, 
presenting a separate class of critical phenomena not related directly to singularities in the bulk free energy.~\cite{Dietrich_DL, Schick, Parry_review} However, the problem of interfacial quantum critical behavior has so far received surprisingly 
little attention,~\cite{Khawaja_99, Anglin_01, Zwerger_11} considering its potential 
practical relevance and the fact that the critical behavior at the boundary is completely distinct to its bulk counterpart. In this paper we adapt an argument due to Cahn,~\cite{Cahn_77} well known in the theory of 
interfacial physics, to argue that the onset of bulk order at zero temperature is accompanied by an interfacial phase transition, which may be of second order even if the bulk transition is discontinuous. 
This surface phase transition can be described by a quantum version of the capillary-wave Hamiltonian. We subsequently apply a functional renormalization-group (RG) technique to analyze the 
interfacial singularities along the zero isotherm, as well as approaching the quantum phase transition from the finite temperature region, and characterize the surface quantum critical regime. 
We assume that the underlying microscopic interactions are short-ranged and the bulk phase transition can be described by an effective bosonic $d$-dimensional field theory characterized by a (bulk) dynamical exponent 
$\tilde{z}$. 
This may apply to systems of static spins \cite{Sachdev_book} as well as itinerant Fermi systems.\cite{Hertz_76, Millis_93} We focus on the case of discrete Ising-type symmetry breaking in the bulk. The generic bulk phase 
diagrams of the systems in question are schematically
 depicted in Fig.~1. The transition can be tuned by varying a non-thermal control parameter $\delta$, or the temperature $T$. At $T=0$ there is a quantum phase transition of second (A) or first (B) order. In the latter case 
a tricritical point generically occurs at $T_{tri}>0$.\cite{Belitz_05_2, Green_05, Misawa_08, Misawa_09, Jakubczyk_09, Jakubczyk_10}

\begin{figure}[h]
\includegraphics[width=8.5cm]{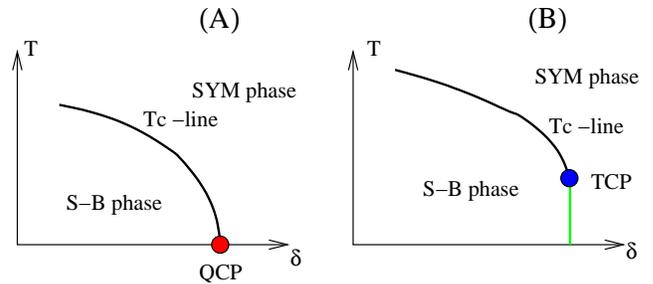}
\caption{A schematic illustration of the bulk phase diagrams. The transition between the symmetric (SYM) and symmetry-broken (S-B) phases can be tuned by varying a nonthermal control parameter $\delta$ 
or the temperature $T$. The second-order phase transition line ($T_c$-line) terminates at a 
quantum critical point (QCP) (A), or becomes first-order below a tricritical temperature $T_{tri}$ (B). The first-order transition line below the tricritical point (TCP) is represented (schematically) 
by the vertical green line.  }  
\end{figure}  
In the present work we consider the system in the symmetry-broken phase ($T<T_c(\delta)$), and subject to an infinitesimal bulk field $h=0^-$ favoring one of the two available bulk phases (phase $\alpha$). 
On the other hand, we impose an external surface field $h_1>0$ acting only at the boundary of the system and giving rise to an opposite ordering tendency (phase $\beta$) at the surface. This implies the formation of a layer of 
the $\beta$-like phase at the wall. The layer may be either microscopic or macroscopic in width, the two situations being separated by what is called an interface unbinding or wetting transition,\cite{Cahn_77, Ebner_77} at which the $\alpha 
- \beta$ interface becomes macroscopically separated from the wall, and the capillary fluctuations diverge.
Such a phenomenological picture may be realized in microscopic models as well as at the level of coarse-grained Landau-type theories supplemented with surface terms.\cite{Binder_DL} Another approach 
employs effective interfacial models, which derive from further coarse-grained Laudau theory.\cite{Buff_65, Lipowsky_DL, Jin_93} The classical versions of all these were extensively studied in course of the last 35 years.    
The quantum variant, relevant for the case of temperatures approaching zero seems however missing. \\
Particular questions addressed in the present paper are: (1) Is the shape of the quantum interface unbinding 
transition line universal in the limit $T\to 0$ like happens to be in the bulk case;\cite{Millis_93, Loehneysen_07, Jakubczyk_08} (2) Is there an analog of the bulk quantum critical regime associated with the quantum interface unbinding transition, 
if so, what are the properties of the system in this regime; (3) The classical interface unbinding transition in $d=3$ is extremely unusual,\cite{Brezin_83, Fisher_85, Lipowsky_87} being characterized by nonuniversal critical exponents and in certain 
cases essentially singular behavior. Does this exotic, nonuniversal behavior persist at $T\to 0$, how is this altered by quantum fluctuations?  

The paper is structured as follows: In Sec.~II we apply a standard argument due to Cahn to argue, that bulk quantum criticality occurring in presence of suitably chosen surface fields   
is generically accompanied by a separate phase transition restricted to the system boundary. In Sec.~III we introduce a phenomenological effective Hamiltonian model capturing the relevant thermal and quantum 
interfacial fluctuations. In Sec.~IV we derive the functional RG flow equation to be solved in Sec.~V at zero temperature and in Sec.~VI in the classical limit (studied already previously). Relying on the 
results of Sec.~V and Sec.~VI we are able to approximately solve the RG flow equation at finite but low temperatures, traversing both the $T=0$ and the classical scaling regimes. This is achieved in Sec.~VII yielding 
results summarized in phase diagrams. We conclude in Sec.~VIII.

\section{Cahn argument}
The original argument due to Cahn \cite{Cahn_77} refers to a situation, where a first order classical phase transition line terminates at a classical critical point. The temperature is increased so that the system approaches 
the critical point, remaining infinitesimally close to the transition in the phase disfavored by the boundary. An interface unbinding transition must occur at the wall before the bulk critical temperature 
is reached. We present the argument in a version adapted to the presently relevant setup. We approach a second-order bulk phase transition (Fig.~1) along any isotherm $T\geq 0$ (case (A)) or $T\geq T_{tri}$ (case (B)). 
The bulk field $h=0^-$ and the surface field is fixed at a positive value $h_1>0$. Assuming no macroscopically wide layer of the $\beta$ phase intrudes the vicinity of the wall, the surface free energy 
density $\sigma_{w\alpha}$ and the free energies (per unit area) associated with the wall-$\beta$ and $\alpha$-$\beta$ interfaces ($\sigma_{w \beta}$ and $\sigma_{\alpha \beta}$ respectively) must fulfill the relation 
\begin{equation}
\sigma_{w\alpha} - \sigma_{w \beta} - \sigma_{\alpha \beta} <0 \;,
\label{Young_ineq} 
\end{equation}
otherwise the system would find it favorable to unbind the $\alpha-\beta$ interface from the wall and form a macroscopically wide layer of the $\beta$ phase. In the vicinity of the critical line however 
$\sigma_{\alpha \beta}\sim (\delta_c-\delta)^{\mu}$ and $|\sigma_{w\alpha}-\sigma_{w\beta}|\sim (\delta_c-\delta)^{\beta_1}$. Here $\beta_1$ is the exponent describing vanishing of the order parameter at the wall as 
the critical point is approached.\cite{Dietrich_DL} The function $\delta_c(T)$ describes the line of critical points and $T_c(\delta)$ is its inverse. 
Plugging in the values relevant for the $d=3$ Ising universality class ($\mu\approx 1.3$, $\beta_1\approx 0.8$),\cite{Dietrich_DL}  
one realizes that condition (\ref{Young_ineq}) is violated sufficiently 
close to the critical line $T_c(\delta)$, implying that at $\delta_w(T)<\delta_c(T)$ the inequality (\ref{Young_ineq}) is replaced with an equality and the interface unbinds. Here $\delta_w(T)$ (or $T_w(\delta)$) 
denotes the interface unbinding transition line. 
The conclusion also holds if mean-field values of the exponents $\mu$ and $\beta_1$ are used, as well as for the case $T=T_{tri}$.\cite{Binder_DL} For this reason the $T_w$-line cannot end at the TCP. For a recent study 
of classical wetting in systems involving a tricritical point see Ref.~\onlinecite{Albano_12}. The argument applies also for $T=0$ in case (A), if the bulk transition at $T=0$ 
falls into the mean-field universality class, which is the usual situation and which we here assume. This also naturally comes out in the calculation in Sec.~V.  
As a consequence, a layer of the $\beta$ phase must be coating the wall for $\delta\in [\delta_w(T), \delta_c(T)[$. In case (B) the argument can be repeated approaching the trictitical point along the triple line and 
 two situations must be distinguished depending on whether the interface is unbound down to $T=0$ or not (see Fig.2). The former offers an interesting route of realizing quantum critical behavior at the boundary 
in the absence of any bulk quantum critical point.   
\begin{figure}[h]
\includegraphics[width=8.5cm]{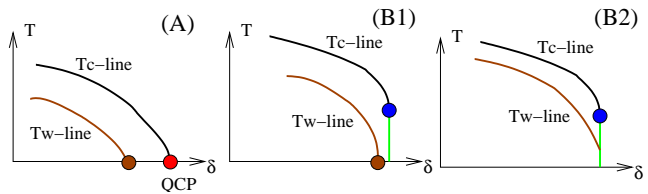}
\caption{A schematic illustration of the phase diagrams as implied by Cahn's argument. A line of interface unbinding transitions necessarily continues to $T=0$ if the bulk phase diagram features a quantum critical 
point (A). In the case of a first-order bulk phase transition below $T_{tri}$ two possibilities occur: The $T_w$-line continues to $T=0$ (B1), or terminates at the finite-$T$ bulk phase transition line (B2).}  
\end{figure}  
As is well known, the interface unbinding transition is accompanied by a divergent correlation length $\xi_\parallel$ controlling the capillary-wave like fluctuations of the $\alpha-\beta$ interface.\cite{Buff_65} 
On the other hand the 
bulk correlation length $\xi_b$ remains microscopic at $\delta_w(T)$.

Let us also note that the Cahn argument breaks down in a number of physically interesting cases involving long-ranged microscopic interactions.\cite{Nightingale_85} However the quantum systems we address here either 
interact via short-ranged forces (static spins) or display efficient screening mechanisms (itinerant fermions). We expect that the Cahn argument is generically valid in the present context. 

\section{Model}
In this section we introduce a phenomenological capillary-wave type model adapted to a situation where quantum-mechanical and thermal fluctuations must be treated on equal footing. The model retains the position and 
imaginary-time dependent location $l({\bf x},\tilde{t})$ of the $\alpha-\beta$ interface above the wall as the fluctuating degree of freedom. It relies therefore on the assumption that all 
the soft modes at the interface unbinding transition are of 
capillary-wave type or that the effect of other soft modes amounts to renormalization of the parameters describing the effective order-parameter action. 
In the classical case this may be proven in $d=2$.\cite{Abraham_80} The quantum effects are incorporated into the model via the imaginary time dependence of the fluctuating degree of freedom.
\cite{Negele_Orland_book} 
In the classical context the capillary-wave Hamiltonian was derived \cite{Jin_93} starting from an underlying bulk Landau theory supplemented by surface terms. In such derivation one fixes the position of the phase 
interface, constraining the accessible order-parameter profiles, and subsequently performs a partial trace respecting this constraint. Since no soft modes responsible for the critical singularities are integrated out in 
this procedure, one may rely on a rather crude approximation retaining only the dominant (saddle-point) term in the constrained trace. For the present case, involving quantum-mechanical effects, such analysis may be 
carried out along fully analogous lines. For systems of static spins the relevant effective bulk theory is a $d+1$-dimensional $\phi^4$ model, where finite-temperature effects are accounted for by a finite-size effect in 
one direction.\cite{Sachdev_book} The analysis of Ref.~\onlinecite{Jin_93} generalizes to the quantum case straightforwardly and yields the interfacial critical exponent $z=1$. The situation is somewhat more complicated 
in Fermi systems, characterized by a Fermi surface and 
soft bulk excitations occurring in the entire thermodynamic phases. This yields the bulk dynamical exponent $\tilde{z}>1$.\cite{Hertz_76} For such cases, in the present study, we shall assume that the damping term in the effective action is 
inherited from the bulk order-parameter theory. We shall therefore neglect the possibility that \textsl{direct} coupling of the fluctuating interface to the soft particle-hole excitations across the Fermi surface changes the way in 
which the imaginary time occurs in the effective action. 
For such cases we shall focus on $z=2$ pertinent to itinerant Fermi systems ordering at wavevector $Q\neq 0$, which may apply either to charge- or spin-density-waves. 

Relying on the above we consider the bare effective action:
\begin{eqnarray}
\mathcal{H}[l(\mathbf{x},\tilde{t})]=\int_0^{1/T} d\tilde{t}\int d^{d-1}x V(l)\nonumber\\
  + T\sum_{\omega}\int\frac{d^{d-1}q}{(2\pi)^{d-1}}[\sigma q^2 + Z_\omega|\omega|^{2/z}]l_{\mathbf{q} \omega} l_{-\mathbf{q} -\omega}\;.
\label{Hamiltonian}
\end{eqnarray}
Here $V(l)$ is a local potential describing interactions between the wall and the $\alpha-\beta$ interface, $\sigma$ is a constant interfacial stiffness coefficient, $l_{\bf{q} \omega}$ is the Fourier transform of 
$l({\bf x}, \tilde{t})$, and $z\in \{1,2\}$. Finally $\omega=2\pi nT$ are the (bosonic) Matsubara frequencies, and the quantum damping term involves a constant $Z_\omega$. 
As already stated, this model may be understood as a coarse-grained Landau-Ginzburg Hamiltonian, out of which the bulk order parameter fluctuations had been 
integrated out, retaining $l(\mathbf{x},\tilde{t})$ as the only degree of freedom. The interfacial potential reads
\begin{equation}
\label{pot}
 V(l)=\tau e^{-l/\xi_b}+ b e^{-2l/\xi_b} + ...\;.
\end{equation}
The parameters $\tau\sim (T-T_w)$ and $b>0$ may (alike in the classical case) be expressed in terms of the underlying Landau-Ginzburg model. For $\tilde{z}=1$ pertinent to systems of static spins the model defined by Eq.~(2) and Eq.~(3) may be derived from the bulk order-parameter action\cite{Sachdev_book} parallelizing Ref.~\onlinecite{Jin_93}. Indeed, the 
quantum effective action is in this case equivalent to the standard classical $\phi^4$-model in dimensionality elevated by 1, and where the system size in one direction is $\beta=1/T$. For this reason it is also very 
plausible that the capillary waves are the only relevant low-energy modes also for the quantum systems. On the other hand, the value of $z$ is usually altered in the cases where the order parameter field interacts with 
a bath of 
generically soft modes like happens to be in the Fermi systems. In such cases one may expect that the value of $z$ depends on the particular bulk instability in question and potentially other characteristics of the system. We have 
performed explicit calculations for $z=1,2$. As the calculation below shows, in $d\geq 3$ the interfacial quantum critical behavior does not show a dependence on the value of $z$ apart from coefficients which are nonuniversal anyway.

We skip higher-order terms and the dependence of $\sigma$ on $l$\cite{Jin_93} in Eq.~(\ref{pot}) At mean-field level the average separation of the interface $l_\pi$ from the wall results from minimizing $V(l)$. 
The interface unbinds as $\tau\sim (\delta-\delta_w(T))\to 0^-$.
 One straightforwardly obtains $l_\pi= \xi_b\log \frac{2b}{-\tau}$ and the interfacial correlation length $\xi_{\parallel}^{-2}\sim V''(l_\pi)=\tau^2/(2b)$. Note that for $\tau>0$ the interface remains unbound and the system 
features soft capillary fluctuations with $\xi_{\parallel}^{-2}=0$. The quantities $\tau$ and $b>0$ can be related to the parameters of a ''semi-microscopic'' Landau-Ginzburg theory. In particular one finds\cite{Jin_93} 
$\tau\sim h_1+ gm_0$, where $g$ is a surface enhancement parameter and $m_0$ the average bulk order parameter.    
 A much more complete description of classical wetting is achieved within nonlocal interfacial models,\cite{Parry_04, Parry_08} where the leading interfacial singularities however turn out to be essentially of the 
same type as those resulting from the much simpler local approach. In Eq.~(\ref{Hamiltonian}) the variables should be understood as coarse-grained over a scale of the order of the intrinsic width of the interface, which in turn is typically of the order of a few 
bulk correlation lengths. The Hamiltonian Eq.~(\ref{Hamiltonian}) is therefore supplemented by an upper momentum cutoff $\Lambda_0$ somewhat lower than $\xi_b^{-1}$. In the practical calculations we will put 
$\Lambda_0=\xi_b^{-1}=1$.

\section{RG flow equations}
Description of the classical interface unbinding transitions is among the early successful applications of functional renormalization group.\cite{fRG_reviews} Within a linearized treatment the critical singularities may 
be captured in 
$d\geq 3$ 
by an analytical description.\cite{Brezin_83, Fisher_85} The present paper may be viewed as an extension of the studies of interface unbinding transitions to the case involving the limit $T\to 0$ and requiring an account 
of quantum fluctuations. 
In what follows we derive 
the relevant RG flow equations from an exact functional RG flow equation for the effective average action $\Gamma_\Lambda [l]$, which is the generating functional for one-particle irreducible vertex functions in the 
presence of an infrared cutoff at scale $\Lambda$. The cutoff parameter $\Lambda\in [0,\Lambda_0]$ and the effective action interpolates between the bare action (Hamiltonian) at $\Lambda\to \Lambda_0$, and the full effective action (Gibbs free energy) as the cutoff is removed 
$\Lambda\to 0$. The exact flow equation \cite{Wetterich_93} reads:
\begin{equation}
\partial_\Lambda \Gamma_\Lambda[l]=\frac{1}{2}\mathrm{Tr}\frac{\partial_\Lambda R_\Lambda (q)}{\Gamma^{(2)}_\Lambda[l]+R_\Lambda (q)}  \;, 
\label{Wetterich_eq}
\end{equation}
where $R_\Lambda(q)$ is a momentum cutoff function added to the inverse propagator to regularize the infrared divergencies, and $\Gamma^{(2)}_\Lambda[l]=\delta^2\Gamma_\Lambda[l]/(\delta l_{\bf{q},\omega}
\delta l_{\bf{-q},-\omega})$. The trace sums over momenta and Matsubara frequencies: $\rm{Tr}=T\sum_{\omega}\int\frac{d^{d-1}q}{(2\pi)^{d-1}}$. Integrating Eq.~(\ref{Wetterich_eq}) with the initial condition at $\Lambda_0$ 
specified by Eq.~(\ref{Hamiltonian}) would yield full solution to the model. Herein we will approximate $\Gamma_\Lambda[l]$ by the following ansatz:
\begin{eqnarray}
 \Gamma_\Lambda[l] = \int_0^{1/T}d\tilde{t} \int d^{d-1}x V_\Lambda (l) \nonumber \\
+ T\sum_{\omega}\int\frac{d^{d-1}q}{(2\pi)^{d-1}}[\sigma q^2 + Z_\omega|\omega|^{2/z}]l_{\bf{q} \omega} l_{-\bf{q} -\omega}\;.
\label{Eff_action}
\end{eqnarray}
This approximation retains functional flow of the local potential $V_\Lambda (l)$ and completely neglects the renormalization of the propagator. The latter does not influence the critical singularities, since the 
anomalous dimension $\eta$ for interface unbinding transitions is robustly zero \cite{Lipowsky_87} and, as we show later on, the precise value of $z$ is of minor importance to the critical behavior. 
The fact that $\eta=0$ makes the present approach easier
 than an analogous treatment of bulk criticality, where anomalous scaling of the propagator requires account of its flow under the RG transformation.\cite{Ballhausen_04, Canet_03, Benitez_09, Benitez_12}   
Evaluation of Eq.~(\ref{Wetterich_eq}) for a 
uniform field $l$ yields the flow of $V(l)$:  
\begin{equation}
\partial_\Lambda V_\Lambda (l) = \frac{1}{2}\mathrm{Tr}\frac{\partial_\Lambda R_\Lambda (q)}{\sigma q^2+Z_\omega |\omega|^{2/z} + V_\Lambda ''(l) + R_\Lambda (q)}\; ,
\label{LPA_flow} 
\end{equation}
where primes denote differentiation with respect to $l$. In the classical limit, where only the contribution from $\omega=0$ is retained, Eq.~(\ref{LPA_flow}) is scale invariant when rewritten in the variables 
\begin{eqnarray}
\label{Class_rescaling}
s=-\log\frac{\Lambda}{\Lambda_0}\;, \;\;\; y=\frac{q^2}{\Lambda^2}\;, \;\;\; \tilde{b}=\sqrt{\sigma}\Lambda^{\frac{3-d}{2}}l\;, \nonumber \\
v_\Lambda (\tilde{b})=V_\Lambda (l)\Lambda^{-d+1}\;, \;\;\; \tilde{r}(y)=\frac{R_\Lambda (q)}{\sigma q^2}\; .
\end{eqnarray}
Namely, we obtain:
\begin{eqnarray}
\label{Class_flow_rescaled}
\partial_s v_\Lambda (\tilde{b})=(d-1)u_\Lambda (\tilde{b}) + \frac{3-d}{2} \tilde{b}\partial_{\tilde{b}} v_\Lambda (\tilde{b}) \nonumber\\
+\frac{S_{d-2}}{4(2\pi)^{d-1}}\int dy y^{\frac{d-3}{2}}\frac{\partial_s r_\Lambda (q)}{1+y^{-1}
\partial^2_{\tilde{b}} v_\Lambda (\tilde{b})+r_\Lambda (y)} \;,
\end{eqnarray}
where we performed the angular integration, and $S_d=2\pi^{(d+1)/2}/\Gamma((d+1)/2)$ is the surface area of a $d$-dimensional sphere of radius 1. In the sharp cutoff limit
\begin{equation}
 \label{Sharp_cutoff}
R_\Lambda (q)=\lim_{\gamma\to\infty}\sigma\Lambda^2\gamma\theta (\Lambda^2-q^2)
\end{equation}
this is equivalent to the continuous limit of the Wilson's approximate recursion relations as derived and analyzed for classical interface unbinding phenomena in Ref.~\onlinecite{Lipowsky_87}.

On the other hand, in the zero-temperature limit $T\to 0$ we have $T\sum_\omega \to \int\frac{d \omega}{2\pi}$ and the scale invariance of Eq.~(\ref{LPA_flow}) is achieved by the transformation Eq.~(\ref{Class_rescaling}) after 
substituting $d\to d+z$. This is a well-known fact, that many quantum-critical systems (at $T=0$) fall into the same universality class as their classical counterparts in dimensionality elevated by the dynamical 
exponent. This is also true for the interfacial phase transitions considered here.

In what follows, we shall rely on a linearized version of Eq.~(\ref{LPA_flow}). Such an approach exploits the fact that close to the transition the effective potential $V(l)$ is very flat except for 
the vicinity of the wall, and the minimum of $V_\Lambda (l)$ is well separated from the system boundary ($l_\pi/\xi_b\gg 1$). The linearized treatment captures the leading classical critical behavior in $d\geq 3$.  
After expanding Eq.~(\ref{LPA_flow}) and dropping a constant term which merely shifts the 
effective potential without influencing its shape, we obtain 
\begin{equation}
\partial_\Lambda V_\Lambda (l) = -\frac{1}{2}\mathrm{Tr}\frac{\partial_\Lambda R_\Lambda (q)}{[\sigma q^2+Z_\omega |\omega|^{2/z}+R_\Lambda (q)]^2}V_\Lambda ''(l)\;.
\label{Eqq}
\end{equation}
We note, that in the classical limit, for $d=3$ and after rescaling according to Eq.~(\ref{Class_rescaling}) this equation is equivalent to the linear RG flow equation analyzed in Ref.~\onlinecite{Fisher_85}. 
This result does not depend on the particular choice of the cutoff function $R_\Lambda (q)$.\cite{Jakubczyk_11}
We rewrite Eq.~(\ref{Eqq}) as
\begin{equation}
 \partial_\Lambda V_\Lambda (l) = -f_T(\Lambda)V_\Lambda ''(l)\;,
\label{diff_eq}
\end{equation}
and split the RHS $f_T(\Lambda)=f_T^{cl}(\Lambda)+f_T^{q}(\Lambda)$ into a classical contribution $f_T^{cl}(\Lambda)$ from $\omega=0$, and a quantum contribution $f_T^{q}(\Lambda)$ from $\omega\neq 0$.

Following, we exploit the freedom of the cutoff choice by taking \cite{Litim_01}
\begin{equation}
 R_\Lambda (q)=\sigma (\Lambda^2-q^2)\theta (\Lambda^2-q^2)\;.
\end{equation}
With this choice the $q$-integration in Eq.~(\ref{diff_eq}) becomes restricted to $q<\Lambda$ and $\sigma q^2 +R_\Lambda (q) =\sigma \Lambda^2$ for $q<\Lambda$. We obtain 
\begin{equation}
 f_T^{cl}(\Lambda)=\frac{S^{d-2}}{(2\pi)^{d-1}}\frac{T\Lambda^{d-4}}{(d-1)\sigma}
\end{equation}
The Matsubara sums in $f_T^{q}(\Lambda)$ are preformed using standard techniques. \\ For $z=1$ we find
\begin{eqnarray}
 f_T^{q}(\Lambda) = \frac{S^{d-2}}{(2\pi)^{d-1}}\frac{T\sigma\Lambda^{d}}{2(d-1)Z_{\omega}^2(2\pi T)^4}\pi^4 \tilde{T}'^4 \times \nonumber\\ \left( -2 + \tilde{T}'^{-1}\coth (\tilde{T}'^{-1}) + \tilde{T}'^{-2} \sinh ^{-2} (\tilde{T}'^{-1}) \right)   \;,
\end{eqnarray}
while for $z=2$ 
\begin{equation}
 f_T^{q}(\Lambda)=\frac{S^{d-2}}{(2\pi)^{d-1}}\frac{T\sigma\Lambda^{d}}{d-1}\frac{2\Psi_1(1+\tilde{T}^{-1})}{(2\pi TZ_\omega)^2}\;,
\end{equation}
where the polygamma functions $\Psi_k(z)$ are defined via $\Psi_{k+1}(z)=\Psi_k'(z)$, $k=0,1,2,...$ and $\Psi_0(z)=\Gamma '(z)/\Gamma (z)$. 
We introduced $\tilde{T}=\frac{2\pi T Z_\omega}{\sigma\Lambda^2}$ and $\tilde{T}' = \sqrt{\frac{Z_\omega}{\sigma}}\frac{2T}{\Lambda}$. For future reference we also define 
$\tilde{T}_z=2\pi^{z-1}\left(\frac{Z_\omega}{\sigma}\right)^{z/2}\frac{T}{\Lambda^z}$.

Since we are interested in the low temperature regime, $\tilde{T}_z$ is small at the beginning of the flow, and the term involving $f_T^{q}(\Lambda)$ dominates. As the cutoff scale is reduced, $\tilde{T}_z$ grows 
and the flow 
may cross over to the classical regime dominated by the term with $f_T^{cl}(\Lambda)$. This will however clearly never happen at $T=0$.

Eq.~(\ref{diff_eq}) can be integrated starting from an initial potential $V_{\Lambda_0}(l)$. This yields 
\begin{equation}
 V_\Lambda (l) = \int dl'(-4\pi F_T(\Lambda))^{-1/2}e^{\frac{(l-l')^2}{4F_T(\Lambda)}}V_{\Lambda_0}(l')\;,
\label{sol_diff_eq}
\end{equation}
where 
\begin{equation}
 -F_T(\Lambda)=\int_\Lambda^{\Lambda_0} d\Lambda ' f_T(\Lambda ')\;.
\end{equation}
We note that although the flow equation was derived from an exact functional flow equation, the problem at hand is absolutely perturbative (at least in $d\geq 3$). Indeed, the same linear flow equation can 
be derived via a somewhat more cumbersome traditional procedure of Wilsonian momentum shell integration. We also note, that the present approximate linear RG approach does not yield any fixed point except 
for the non-interacting.\cite{Lipowsky_87} The procedure to be applied in Sec.VII in order to compute the phase diagrams combines those of Ref.~\onlinecite{Millis_93} and Ref.~\onlinecite{Fisher_85}. 
We renormalize to a scale at which the curvature of the renormalized, 
rescaled potential becomes of the order unity (in units where the microscopic scale $\Lambda_0^{-1}=1$). 
We then treat the renormalized potential with a saddle-point approximation and rescale back to the physical lengths. The flow is split into a quantum ($\tilde{T}_z<1$) and classical ($\tilde{T}_z>1$) stages. We scale 
the variables according to Eq.~(\ref{Class_rescaling}) in the classical regime, and replace $d$ with $d+z$ in Eq.~(\ref{Class_rescaling}) for the flow in the quantum regime.

\section{Solution at zero temperature}
We present details of the solution for $z=2$. The procedure for $z=1$ is fully analogous. At $T=0$ the classical term $f_T^{cl}(\Lambda)=0$ and $\tilde{T}\ll1$ at any $\Lambda$. By expanding the polygamma function around $\tilde{T}=0$: $\Psi_1(1+\tilde{T}^{-1})=\tilde{T}+\mathcal{O}(\tilde{T}^2)$ we find 
\begin{equation}
 f_0^q(\Lambda) = \lim_{T\to 0} f_T^q(\Lambda)=C^d\Lambda^{d-2}\;, 
\label{Eqq1}
\end{equation}
where we introduced $C^d=\frac{2S^{d-2}}{Z_\omega (d-1) (2\pi)^d}$.
The integral defining $F_T(\Lambda)$ is now straightforward and one finds
\begin{equation}
V_\Lambda (l) = \int dl' (4\pi K(\Lambda))^{-1/2}e^{-(l-l')^2/4K(\Lambda)} V_{\Lambda_0}(l')\;,
\label{Ren_V_T0}
\end{equation}
where $K(\Lambda)=\frac{C^d}{d-1}(1-\Lambda^{d-1})$. This implies that the effect of fluctuations at any $d>1$ is merely to smear the bare potential by convoluting it with a Gaussian of finite width. This is 
reminiscent of the solution to the classical problem in $d>3$ \cite{Fisher_85} and consistent with the expectation that the quantum interface unbinding transition at $T=0$ is in the same universality class as its 
classical counterpart in 
higher dimensionality $D=d+z$.\cite{Sachdev_book} Note that the transition occurs at $\tau=0$ and the critical value of $\delta_w (T=0)$ is not affected by fluctuations. In the limit $\tau\to 0^-$ ($\delta\to \delta_w (T=0)^-$) 
one finds $\xi_{\parallel}^{-1}\sim (\delta_w (T=0)- \delta)$ and a logarithmic divergence of $l_\pi$.\\
The solution proceeds along the same line for $z=1$, where in Eq.~(\ref{Eqq1}) $\Lambda^{d-2}$ is replaced by $\Lambda^{d-3}$ and the constant $C^d$ takes a different form. Again, the renormalized potential $V_\Lambda (l)$ is nothing more 
than a  convolution of the bare one $V_{\Lambda_0}(l)$ with a Gaussian whose width is of microscopic extent and the critical properties are described by the mean field theory (MFT). The general solution for $z\in\{1,2\}$ 
can be written as
\begin{equation}
V_\Lambda (l) = \int dl'\frac{1}{\sqrt{\pi\Delta_\Lambda^z}}e^{-(l-l')^2/\Delta_\Lambda^z}V_{\Lambda_0}(l')\;, 
\end{equation}
where 
\begin{equation}
 \Delta_\Lambda^z = C_z^d\left(1-\Lambda^{d+z-3}\right)\;,
\end{equation}
with $C_1^d=\frac{S^{d-2}}{(2\pi)^{d-1}\sqrt{Z_\omega \sigma}(d-1)(d-2)}$, and $C_2^d=\frac{8S^{d-2}}{(2\pi)^dZ_\omega (d-1)^2}$.

\section{Solution in the classical limit}
As we already mentioned, in the classical limit the quantum contribution to the flow equation vanishes ($f_T^q(\Lambda)\to 0$), the flow is dominated by the classical term, 
and up to a multiplicative constant $A(d)$ the linearized flow equation is equivalent to the equation analyzed in Ref.~\onlinecite{Fisher_85}. Occurrence of $A(d)$ is due to the cutoff choice, which is different here 
as compared to Ref.~\onlinecite{Fisher_85}. 
Importantly $A(3)=1$,\cite{Jakubczyk_11} which holds irrespective of $R_\Lambda (q)$. On the other hand, above the upper critical dimension $d>3$, the quantity $A(d)$ has no impact on the critical behavior. 
Our results in the high temperature limit coincide with Ref.~\onlinecite{Fisher_85}, and we quote them here. The bare effective potential 
\begin{equation}
 V_{\Lambda_0}(l)= A_{\Lambda_0}(l)+R_{\Lambda_0}(l)+W_{\Lambda_0}(l)
\label{bare_V_split}
\end{equation}
is split into an attractive tail $A_{\Lambda_0}(l)=\tau e^{-l/\xi_b}\theta(l)$, a repulsive tail $R_{\Lambda_0}(l)=be^{-2l/\xi_b}\theta(l)$ and a wall repulsion contribution $W_{\Lambda_0}(l)$. Following 
Ref.~\onlinecite{Fisher_85} we model 
the latter by a soft-wall potential $W_{\Lambda_0}(l)=w\theta (-l)$, $w$ being a constant. For $d>3$ fluctuations do not introduce significant renormalization to the bare potential, and the renormalization amounts to
convoluting the bare potential with a Gaussian of finite width. The critical singularities are identical to those of MFT. The situation is quite distinct in $d=3$, 
where the convolution width becomes infinite. In particular, for a range of parameters, the wall repulsion renormalization becomes essential for the critical behavior.    
The three contributions to the renormalized potential from the different interaction terms 
in Eq.~(\ref{bare_V_split}) must be carefully analyzed at the scale $\Lambda^*$, where the curvature at the minimum is of the order unity. Applying a mean-field approximation at this scale and returning to the physical length 
units yields the critical behavior. The interplay between the three terms in Eq.~(\ref{bare_V_split}) leads to three fluctuation regimes showing non-universal critical behavior determined by the dimensionless parameter 
\begin{equation}
 \omega_w=\frac{T_w}{4\pi\sigma}
\end{equation}
For the capillary correlation length exponent $\nu_{\parallel}$ defined by $\xi_{\parallel}\sim (\delta_w-\delta)^{-\nu_\parallel}$ one finds 
\begin{equation*}
\nu_\parallel =
\begin{cases}
(1-\omega_w)^{-1} & \text{for  } \omega_w \in (0,1/2)\\
(\sqrt{2}-\sqrt{\omega_w})^{-2} & \text{for } \omega_w \in (1/2,2) \\
\infty & \text{for } \omega_w \in (2,\infty).
\end{cases}
\end{equation*}

For $\omega_w\to 2^-$ the exponent $\nu_\parallel$ diverges and one finds an essential singularity, whereby $\xi_{\parallel}\sim e^{\alpha/(\delta_w-\delta)}$. The film thickness diverges logarithmically for 
$\omega_w<2$ and $l_\pi \sim 1/(\delta_w-\delta)$ for $\omega_w>2$. Remarkably, the critical value $\delta_w$ is not shifted by fluctuations for $\omega_w<2$. One of the major questions raised in this paper concerns the 
fate of this extremely unusual picture in the quantum limit. 
   
\section{Computation of the phase diagram}

\begin{figure}[h]
\includegraphics[width=8.5cm]{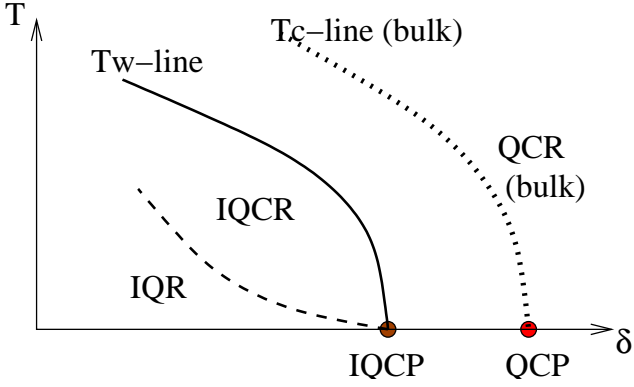}
\caption{A schematic illustration of the computed interfacial phase diagram for case (A) (see Fig.2). The interfacial part of the diagram is the same for the case (B1). The interface unbinding transition 
line $T_w$ is 
parallel to the bulk $T_c$-line and terminates at $T=0$ with an interfacial quantum critical point (IQCP). The power law describing the shape of the $T_w$-line is the same as the for the $T_c$-line and is 
determined by the 
bulk properties (dimensionality and the bulk dynamical exponent). The interfacial correlation length $\xi_{\parallel}$ is infinite in the region between the $T_w$ and the $T_c$ lines. 
A crossover line separates the interfacial quantum regime (IQR) from the quantum critical regime (IQCR). The interfacial dynamical exponent $z$ determines the shape of this crossover line, which is straight for $z=1$ and 
parabolic for $z=2$. 
  In both regimes and for all $d\geq 3$ the interfacial height $l_\pi$ diverges logarithmically upon 
approaching the IQCP. The behavior of the correlation length $\xi_{\parallel}$ is always mean-field in $d>3$. In $d=3$ it is mean-field in the IQR regime, and $\xi_{\parallel}\sim T^{1/(1-D_3T)}$ when the IQCP is 
approached along a straight line from the IQCR regime.}  
\label{Phase_diag_result}
\end{figure}

We  now present a calculation of the interfacial phase diagram at $T \geq 0$. We integrate the linearized RG flow Eq.~(\ref{diff_eq}) at finite, low temperature (such that $\tilde{T}_z\ll 1 $ at $\Lambda=\Lambda_0$). 
Eq.~(\ref{bare_V_split}) acts as
the initial condition. The 
initial stage of the flow is dominated by the quantum contribution and we expand Eq.~(\ref{diff_eq}) around the $\tilde{T}\to 0$ limit, and rescale according to the zero-temperature dimensions, where 
$r=\Lambda^{\frac{3-d-z}{2}}l$, $u_\Lambda (r)=V_\Lambda (l) \Lambda^{-d-z+1}$. For simplicity in the definition of $r$ we dropped the $\sqrt{\sigma}$ factor, which in the present context is just a scale-independent 
constant anyway. Following the flow of the 
renormalized, rescaled potential, we identify two regimes in the space spanned by $(\delta, T)$. In the first case, the parameter $\tilde{T}_z<1$ at the scale $\Lambda^*$, where the potential curvature at the minimum 
becomes of order unity. We 
terminate the flow at $\Lambda^{*}$ and, since the scale where $f_T^{cl}(\Lambda)$ becomes significant was never reached, conclude that the system properties are essentially the same as in the $T=0$ case. This 
regime is an analog to the quantum (or $T=0$) regime in the case of the well-developed theory of bulk quantum phase transitions. The other regime occurs if the scale $\Lambda^*$ is not reached before $\tilde{T}_z\approx1$, 
whereby $\Lambda=\Lambda_{cr}
$. At $\Lambda < \Lambda_{cr}$ the flow becomes dominated by the classical term $f_T^{cl}(\Lambda)$. We now neglect the quantum contribution 
$f_T^{q}(\Lambda)$, and perform the classical rescaling. The flow equations become equivalent to those occurring in the high-temperature limit, where $\Lambda_{cl}$ acts as the upper cutoff and the initial condition is 
provided by $V_{\Lambda_{cr}}(l)$ obtained by the first stage of the flow integration. The (approximate) matching procedure at $\Lambda_{cr}$ is analogous to that of Ref.~\onlinecite{Millis_93}, and the regime $\Lambda <\Lambda_{cr}$ is studied along the 
same lines as Ref.~\onlinecite{Fisher_85}. We give details of the calculation below, and the results are summarized in the phase diagram Fig. \ref{Phase_diag_result}.

\subsection{Flow for $\Lambda>\Lambda_{cr}$ }
According to the procedure outlined above, we calculate the contributions to the renormalized potential $V_\Lambda (l)$ from the three terms in Eq.~(\ref{bare_V_split}). We put $\Lambda_0=\xi_b^{-1}=1$ and consider 
$\tau<0$. We obtain:
\begin{equation}
 A_\Lambda (l)=\frac{1}{2}\tau e^{\sqrt{\Delta_\Lambda^z/(16\pi)}-l}\mathrm{Erfc}\left[\frac{\Delta_\Lambda^z-2l}{2\sqrt{\Delta_\Lambda^z}}\right]\;,
\end{equation}
\begin{equation}
 R_\Lambda(l)=\frac{1}{2}b e^{\sqrt{\Delta_\Lambda^z/\pi}-2l}\mathrm{Erfc}\left[\frac{\Delta_\Lambda^z-l}{\sqrt{\Delta_\Lambda^z}}\right]\;,
\end{equation}
\begin{equation}
W_\Lambda (l)=\frac{1}{2}w\mathrm{Erfc}\left[\frac{l}{\sqrt{\Delta_\Lambda^z}}\right]\;.
\end{equation}
We observe that $\Delta_\Lambda^z$ remains small and only the terms involving $l$ in the arguments of the error function contribute to the renormalized potential at large $l$. 
The term $W_\Lambda (l)$ decays as $e^{-l^2}$ at $l\gg 1$, and rapidly approaches $w$ for negative $l$. Expanding the error functions yields:
\begin{equation}
 A_\Lambda(l) \approx \tau e^{\sqrt{\Delta_\Lambda^z/16\pi}-l} \theta(l) \;,
\end{equation}
\begin{equation}
 R_\Lambda(l) \approx b e^{\sqrt{\Delta_\Lambda^z/\pi}-2l} \theta (l) \;,
\end{equation}
\begin{equation}
 W_\Lambda(l) \approx w\theta(-l) \;.
\end{equation}
We neglected irrelevant Gaussian tails.

 We now rescale according to $r=\Lambda^{\frac{3-d-z}{2}}l$, $v_\Lambda (r)=V_\Lambda (l) \Lambda^{-d-z+1}$ and analyze the minimum $r_\pi$ of the rescaled flowing potential $v_\Lambda (r)$, 
and the curvature $v_\Lambda '' (r_\pi)$. We consider two scales: $\Lambda_{cr}$, where the parameter $\tilde{T}_z=1$ and the flow crosses over to the classical scaling regime; and $\Lambda^*$, where $v_\Lambda ''(r_\pi)=1$. 
If $\Lambda^*>\Lambda_{cr}$ classical scaling is never reached and the critical behavior is the same as for $T=0$. Solving the inequality 
\begin{equation}
 \Lambda^*>\Lambda_{cr}
\end{equation}
we find
\begin{equation}
T<(-\tau)^z 2\pi^{z-1}\left(\frac{Z_\omega}{\sigma}\right)^{z/2}b^{-z/2}e^{z\sqrt{C_z^d}/(4\sqrt{\pi})}\;.
\label{cross_line} 
\end{equation}
The central observation here is that the crossover line approaches $\tau=0$ linearly for $z=1$ and parabolically for $z=2$. The result given by Eq.~(\ref{cross_line}) is valid in any dimensionality $d\geq 2$. 
Apart from the constant encoded in $C_z^d$ the shape of the crossover line does not depend on the dimensionality $d$ (which is also the case of the corresponding bulk system in $d+\tilde{z}\geq 4$).

%
For the opposite case, where Eq.~(\ref{cross_line}) does not hold, the flow reaches the scale $\Lambda_{cr}$ and crosses over to the classical scaling regime.  

\subsection{Flow for $\Lambda <\Lambda_{cr}$ }
In the classical scaling regime $\Lambda <\Lambda_{cr}$ the flow becomes dominated by the classical term $f_T^{cl}(\Lambda)$ in Eq.~(\ref{diff_eq}). The quantum-mechanical effects are encoded in the potential 
$V_{\Lambda_{cr}}(l)$, which was obtained by integrating the quantum regime of the flow, and which now serves as the initial condition. Therefore both quantum and thermal fluctuations may now influence the 
system properties. We consider the solution Eq.~(\ref{sol_diff_eq}) to the linear flow equation Eq.~(\ref{diff_eq}) treating $\Lambda_{cr}$ as the upper scale and $V_{\Lambda_{cr}}(l)$ as computed above as the 
initial condition. We neglect the term $f_T^{q}(\Lambda)$, rescale according to Eq.~(\ref{Class_rescaling}), and terminate the flow at the scale $\Lambda^*$, when the curvature of the rescaled, renormalized potential 
at the minimum becomes of order unity. Since the correlation length  at this scale is one, the physical correlation length $\xi_{\parallel}$ is of order $(\Lambda^*)^{-1}$. We extract 
information on the critical behavior of the interfacial height $l_\pi$ analogously. \\
From the technical point of view this part of the analysis parallels Ref.~\onlinecite{Fisher_85}. The differences amount to treating the temperature-dependent scale $\Lambda_{cr}$ as the upper cutoff and the potential computed in the 
preceding section and evaluated at $\Lambda_{cr}$ as the initial condition. 
For the present case 
\begin{equation*}
-F_T(\Lambda) =
\begin{cases}
\frac{D_d T}{d-3}\left(\Lambda_{cr}^{d-3}-\Lambda^{d-3}\right) & \text{for  } d>3\\
D_d T\log\frac{\Lambda_{cr}}{\Lambda} & \text{for } d=3 \; ,
\end{cases}
\end{equation*}
with 
\begin{equation}
D_d=\frac{S^{d-2}}{(2\pi)^{d-1}(d-1)\sigma}\;. 
\end {equation}
The convolution width as given by $-F_T(\Lambda)$ is obviously finite for any $T>0$ for $d>3$ and the resulting critical behavior is purely mean-field. The situation is 
somewhat more interesting for $d=3$. Parallelizing Ref.~\onlinecite{Fisher_85} we find that at low $T$ the system will find itself in a regime analogous to the classical $\omega_w<1/2$ case, where $\omega_w\sim T$. This 
happens because a temperature factor suppresses a quantity analogous to the capillary parameter $\omega_w$ in the present calculation.  We find:
\begin{equation}
- \log \Lambda^*= \log \left(\frac{2b'}{\tau'^2}\right)^{1/(2-2D_3T)}+\frac{2D_3T\log \Lambda_{cr}}{2-2D_3T}\;,
\end{equation}
 where $\tau'=\tau e^{\sqrt{\Delta_\Lambda^z/16\pi}}$, $b'=b e^{\sqrt{\Delta_\Lambda^z/\pi}}$. 
It follows that 
\begin{displaymath}
 \xi_{\parallel}\sim\left(\frac{\sqrt{2b'}}{-\tau'}\right)^{1/(1-D_3T)} = 
\end{displaymath}
\begin{displaymath} 
\left(\frac{\sqrt{2b}}{-\tau}e^{\sqrt{C_z^3/16\pi}}\right)^{1/(1-D_3T)} \times
\end{displaymath}
\begin{equation}
\left(1-\sqrt{C_z^3/(64\pi)}2\pi^{z-1}(Z_\omega/\sigma)^{z/2}  T\right)^{1/(1-D_3T)}\;.
\end{equation}

The equilibrium interfacial width is given by 
\begin{equation}
 l_{\pi}=-\log\Lambda^* (1+2D_3T)\;.
\end{equation}
Approaching the interfacial quantum critical point along a straight line we put $\tau \sim -\gamma T$, $\gamma>0$.
The conclusions of the calculation in $d=3$ are summarized as follows:\\
1) The leading singular behavior of the correlation length is of the type $\xi_{\parallel}\sim T^{-1/(1-D_3T)}$ implying that the temperature dependence appears in the exponent. As $T\to 0$, the exponent approaches the 
MF value $-1$. \\
2) The value of the dynamical exponent does not influence either the leading singularity or the leading correction (apart from a coefficient). This will not be true for $d<3$. \\
3) The divergence of the interfacial width in the quantum limit $T\to 0$ is always logarithmic. Temperature dependencies due to quantum effects merely renormalize the proportionality coefficient. \\
The calculation shows that the critical behavior occurring in $d=3$ when we approach the interfacial quantum critical point from the interfacial quantum critical regime is equivalent to the case of classical 
transition at $T\to 0$. This situation is unlike the corresponding bulk transition and is hard to predict without performing the calculation. Again, we do not expect this equivalence to persist for $d<3$. 

Neither the model defined by Eq.~(2) and Eq.~(3) nor the above RG analysis require that the bulk transition is second order at $T=0$. Therefore our results are valid also for the case B1 in Fig.~2. Case B2 displays no 
interfacial QCP and is not covered by the present caluclation.   

\subsection{Interfacial phase transition line}
As we have observed, the interfacial transition temperature is not renormalized by fluctuations in any of the cases studied here. The shape of the $T_w$-line is therefore fully determined by the intrinsic dependence 
of the parameter $\tau$ on temperature. Invoking Ref.~\onlinecite{Jin_93} we find that $\tau\sim h_1+g m_0$. When the temperature is varied, the bulk order parameter $m_0\sim (T_c-T)^\beta$. This implies 
$\tau\sim(\delta-\delta_w(T))=C_1+C_2 (\delta-\delta_c(T))^\beta$. For $\delta=\delta_w(T)$ we find $\delta_c(T)-\delta_w(T)=const$. For $d > 2$ one has $T_c\sim (\delta-\delta_0)^\psi$, where the shift exponent 
$\psi=\tilde{z}/(d+\tilde{z}-2)$.\cite{Millis_93} The shape of the $T_w$-line follows the same power law as the $T_c$-line in $d\geq 3$. 

We have however nothing to say about the shape of the $T_w$-line in cases B1 and B2 (Fig.~2). Alike the first-order bulk transition lines, these curves are not expected to display universal behavior.

\section{Summary}
We have analyzed interfacial phenomena occurring in presence of bulk quantum phase transitions. Applying a reasoning adapted from the theory of classical interfacial phase transitions we have argued that interfacial unbinding 
transitions generically accompany quantum critical phenomena. Notably, the boundary transitions may occur (and be of second order) even if the bulk transitions are discontinuous. An interface unbinding 
transition implies a divergent interfacial length scale in a portion of the phase diagram within the bulk-ordered regime. This in turn gives rise to two scaling regimes in a region where the bulk is ordered but the 
interface is bound. We have found that, unlike the case of bulk transitions, the interfacial transition line in the ($\delta$, $T$) plane is not shifted by fluctuations, and (for the case of a continuous transition 
in the bulk) is determined by the shape of the bulk $T_c$-line. Relying on a quantum version of an effective interfacial Hamiltonian, we have computed the crossover between the quantum ($T=0$) and the quantum-critical 
interfacial regimes. Our result for the shape of the crossover line should be valid in arbitrary dimensionality $d\geq 2$. The critical behavior in the quantum regime is purely mean-field for $d>2$, as anticipated from 
quantum to classical mapping arguments. So is the system's behavior in the interfacial quantum-critical regime for $d>3$. Our calculations in the quantum critical regime are reliable only for $d\geq 3$. In the $d=3$ case the 
interfacial correlation length diverges with a temperature dependent critical exponent, where the temperature-dependent part asymptotically vanishes. This feature is closely related to the nonuniversality of the 
classical critical wetting in $d=3$. The crossover between the two interfacial scaling regimes is rather weak and in fact detectable only via the temperature-dependence of the interfacial correlation length exponent 
$\nu_{\parallel}$. In addition to the interfacial correlation length, we have computed the critical behavior of the average 
interfacial width, and found logarithmic singularities in both scaling regimes. We anticipate that the crossover between the two regimes is much more pronounced for $d<3$. In particular, for $d=2$ and $z=1$ the system 
at $T=0$ may be in the same universality class as classical wetting in $d=3$ yielding a very reach spectrum of possible critical behavior. On the other hand, the classical interface unbinding transition in $d=2$ is governed by 
non-mean-field universal exponents. The properties of the quantum-critical regime in $d=2$ is an interesting 
problem, beyond the reach of the analytical linearized functional RG methodology, but open to be addressed within a numerical approach in near future.\\

\begin{acknowledgments}
I would like to thank J. Bauer, B. Delamotte, N. Dupuis, B. Halperin, N. Hasselmann, C. Husemann, M. Napiorkowski, A. Maciolek, W. Metzner, S. Sachdev, P. Strack and H. Yamase for very useful discussions. I also 
acknowledge the hospitality of the Max Planck Institute for Solid Research in Stuttgart and Pierre et Marie Curie University in Paris where substantial parts of this work were carried out. 
\end{acknowledgments}

\end{document}